\documentstyle[11pt]{article}
\catcode`\@=11
\begin{document} \openup6pt

\title{ Holographic Dark Energy Model with
Modified  Variable Chaplygin Gas }

\author{B. C. Paul\thanks{Electronic mail : bcpaul@iucaa.ernet.in}  \\
    Physics Department, North Bengal University, \\
Siliguri, Dist. : Darjeeling, Pin : 734 013, West Bengal, India 
}

\date{}

\maketitle

\vspace{0.5in}

.\begin{abstract}

In this letter we consider  a correspondence between holographic dark energy and variable   modified Chaplygin gas to obtain a holographic dark energy model of the universe. The corresponding  potential  of the scalar field has been reconstructed which describes the variable Chaplygin gas. The stability of the holographic dark energy in this case is also discussed.

\end{abstract}

\vspace{0.2cm}


\vspace{4.5cm}

\pagebreak

\section{ Introduction:}

Recent cosmological observations namely, high redshift surveys of type Ia Supernovae, WMAP, CMB etc. predict that the present universe is passing through an accelerating phase of expansion [1]. It has also been predicted from COBE that  the present universe might have  emerged from an inflationary phase in the early universe.  The above two observational facts in the universe do not find explanation in the framework of Einstein general theory of Relativity (GTR) with normal matter. It is known that  early inflation may be realized in a  semiclassical theory of gravity where matter is described by quantum fields [2]. Starobinsky obtained inflationary  solution considering a curvature squared term in the Einstein-Hilbert action [3] long before the advent of inflation was known. However, the efficacy of inflation is known only after the seminal work of Guth who first employed the phase transition mechanism to accommodate inflation. Thereafter more than  a dozen of inflationary model came up. In some of these inflation is realized by  (i)  a modification of  gravitational sector of action introducung higher order terms  or [4], (ii) a modification of the matter sector  taking into account scalar fields.
In the later case inflation is obtained with an  equation of state $p = \omega \; \rho$, if $\omega = -1$, but  the present accelerating phase can be realized with $\omega < -1$. The usual fields in the standard model of particle physics are not suitable to obtain such accelerating phase of the universe. In fact it is a challenge to theoretical physics to describe the origin of such matter density. Recent astronomical observations predict that dark energy content of the universe is about 76 \% of the total energy density of the universe. To accommodate such a huge energy  density various kinds of exotic matters are considered as  possible candidate for the dark energy. Chaplygin gas is one such candidate  with  an equation of state (EOS)  $p = - \frac{B}{\rho}$ [5], where $\rho$ and $p$ are the energy density and pressure respectively and $B$ is a constant. Subsequently, a modified  form of the equation of state $p = - \frac{B}{\rho^{\alpha}}$ with $0 \leq \alpha \leq 1$ was also considered to construct a viable cosmological model
[6, 7], which is  known as generalized Chaplygin gas (GCG). It has two free parameters $B$ and $\alpha$ respectively.  The EOS for GCG has been further modified to   $p = A \rho - \frac{B}{\rho^{\alpha}} \; $, with $0 \leq \alpha \leq 1$ which is known as modified GCG [8, 9]. In a flat Friedmann model it is shown [6]  that the modified generalized Chaplygin gas may be equivalently described in terms of a homogeneous minimally coupled scalar field $\phi$. Barrow [10] has outlined a method  to fit Chaplygin gas in FRW universe. Gorini {\it  et al.} [11] using the above scheme obtained
the corresponding homogeneous scalar field $\phi (t)$  in a  potential $V(\phi)$ which can be used to obtain a viable cosmological model with modified Chaplygin gas. Another form of EOS for Chaplygin gas [12] is considered recently  which is given by
\begin{equation}
p = A \rho - \frac{B(a)}{\rho^{\alpha}} \; \;  \; with \; \; 0 \leq
\alpha \leq 1,  
\end{equation}
with a variable $B = B_{o} a^{- 3n}$, $B_o$ is a constant and $a$ is the scale factor of the universe. Guo and Zhang [13] obtained cosmological model using the EOS for variable Chaplygin gas. Cosmological observational results are used to study the constraints [14] 
\\

Recently holographic principle [15, 16] is incorporated in cosmology [17-20] to track the dark energy content of the universe following the work of Cohen {\it et al.} [21]. Holographic principle is a speculative conjecture about quantum gravity theories proposed by G't Hooft. The idea is subsequently promoted by Fischler and Susskind [15] claiming that all the information contained in a spatial volume may be represented by a theory that lives on the boundary of that space.
For a given finite region of space it may  contain matter and energy within it. If this energy be less than  a critical value then the region collapses to a black hole. As a  black hole is known theoretically to have an entropy which is proportional to its surface area of its event horizon. A black hole event horizon encloses a volume, thus a  more massive  black hole have larger event horizon and encloses larger volume. The most massive black hole that can fit in a given region is the one whose event horizon corresponds exactly to the boundary of the given region under consideration. The maximal limit of entropy for an ordinary region of space is directly proportional to the surface area of the region and not to its volume. Thus, according to holographic principle, under  suitable conditions all the information about a physical system inside a spatial region is encoded in the boundary. The basic idea of a holographic dark energy in cosmology is that the saturation of the entropy bound may be related to an  unknown ultra-violet (UV)  scale $\Lambda$ to some known comological scale
in order to enable it to find a viable formula for the dark energy which may be quantum gravity in origin and it is  characterized by $\Lambda$. The choice of UV-Infra Red (IR) connection from the covariant entropy bound leads to a universe dominated by blackhole states. According to   Cohen {\it et al.} [21] for   any state in the Hilbert space with energy $E$, the  corresponding  Schwarzschild radius $R_s \sim E$, may be less than the IR cut off value $L$ (where $L$ is a cosmological scale). It is possible to derive a relation between the UV  cutoff $\rho_{\Lambda}^{1/4}$ and the IR cutoff which eventually leads to a constraint   $ \left( \frac{8 \pi G}{c^2} \right) L^3  \left( \frac{\rho_{\Lambda}}{3}\right) \leq L$ [22] where $\rho_{\Lambda}$ is the energy density corresponding to dark energy characterized by $  \Lambda$, $G$ is Newton's gravitational constant  and $c$ is a parameter in the theory. The holographic dark energy density  is
\begin{equation}
\rho_{\Lambda} = 3 c^2 M_P^2 L^{-2},
\end{equation}
where $M_P^{-2} = 8 \pi G$.
It is known that the present acceleration may be described if $\omega_{\Lambda} =\frac{p_{\Lambda}}{\rho_{\Lambda}} < - \frac{1}{3}$. If one considers $L \sim \frac{1}{H}$ it gives  $\omega_{\Lambda} =0$.
A holographic cosmological constant model based on Hubble scale as IR cut off does not permit accelerating universe.
It is also  examined [17]  that the holographic dark energy model based on  the particle horizon as the IR cutoff is not suitable for an accelerating universe. However, an alternative model of dark energy  using particle horizon in closed model is also proposed [23].
 Li [18] has obtained an accelerating universe considering  event horizon  as the cosmological scale. The model is consistent with the cosmological observations.  Thus to have a model consistent with observed universe one should adopt the covariant entropy bound and choose $L$ to be event horizon [24].  Considering  a correspondence of  holographic dark energy and  Chaplygin gas the field potential is reconstructed [25, 26]. 
In this paper  we consider EOS given by  (1) and set up a correspondence with holographic dark energy to reconstruct scalar field potential.

The paper is organized as follows : in sec. 2, the relevant field
equation with  modified  variable Chaplygin gas in FRW universe is presented. Considering correspondence of holographic
dark energy fields with  modified variable chaplygin gas, we determine the field  and  the corresponding potential is reconstructed.  in sec. 3,
squared speed of sound for holographic dark energy is evaluated for  stability analysis. Finally in sec. 4,  a brief
discussion is given.

\section{ Field Equation and  Modified Variable Chaplygin Gas :}

The Einstein's field equation is
given by
\begin{equation}
R_{\mu \nu} -\frac{1}{2} g_{\mu \nu} R =  8 \pi G  \; T_{\mu \nu}
\end{equation}
where $T_{\mu \nu}$ is the energy
momentum tensor.

We consider a Robertson-Walker (RW)  metric given  by
\begin{equation}
ds^{2} = - dt^{2} + a^{2}(t) \left[ \frac{dr^{2}}{1- k r^2} + r^2 ( d\theta^{2} + sin^{2} \theta \;
d  \phi^{2} ) \right]
\end{equation}
where $a(t)$ is the scale factor of the universe. The energy momentum tensor is  $T^{\mu}_{\nu} = ( \rho, p,
p, p)$ where  $\rho$ and $p$ are energy density and pressure
respectively.

Using RW metric (4) and  the energy momentum tensor, the  Einstein's field equation (3)  yields
\begin{equation}
H^{2}+\frac{k}{a^2} =  \frac{1}{ 3 M_{P}^2 } \rho
\end{equation}
where we use $ 8 \pi G = M_{P}^{2}$. The conservation equation for
matter is given by
\begin{equation}
\frac{d\rho}{dt} + 3 H (\rho + p) = 0 
\end{equation}
where $\rho = \rho_{matter} + \rho_{\Lambda}$. For  modified variable chaplygin gas (henceforth, VCG), we use EOS given  by eq. (2) in  eq. (6), which yields
\begin{equation}
\rho_{\Lambda} = \left(  \frac{(1+\alpha) B_o}{(1+\alpha)(1+A) -n} \frac{1}{a^{3n}} + \frac{C_o}{a^m} \right)^{\frac{1}{1+
\alpha}}
\end{equation}
where $C_o$ is an integration constant and we denote
$ m = 3(1 + A) ( 1 + \alpha)$ . We now define the following
\begin{equation}
\Omega_{\Lambda} = \frac{\rho_{\Lambda}}{\rho_{cr}}, \; \Omega_{m} = \frac{\rho_{m}}{\rho_{cr}}, \;
\Omega_{k} = \frac{k}{a^2 H^2}
\end{equation}
where $\rho_{cr} = 3 M_P^2 H^2$, $\Omega_{\Lambda}$, $ \Omega_m $
and $\Omega_k$ represent  density parameter corresponding to
$\Lambda$, matter and curvature respectively.

We assume  here that the origin of dark energy is a scalar field. Making use of  Barrow's scheme [10], we get the following
\begin{equation}
\rho_{\phi} = \frac{1}{2} \dot{\phi}^2 + V (\phi) = \rho = \left(  \frac{
B_1}{a^{3n}}+  \frac{C_o}{a^m} \right)^{\frac{1}{\alpha +1}},
\end{equation}
\begin{equation}
p_{\phi} = \frac{1}{2} \dot{\phi}^2 - V (\phi) = p =  \frac{  
\frac{(n-1-\alpha) B_1}{(1+\alpha) a^{3n}}
    + \frac{AC_o}{a^m}}{  \left(
\frac{B_1}{a^{3n}}+ \frac{C_o}{a^m} \right)^{\frac{\alpha}{1+\alpha}}},
\end{equation}
where $B_1 = \frac{(1+\alpha) B_O}{(1+\alpha)(1+A) -n}$.
Now the corresponding scalar field potential and its
kinetic energy  term is obtained from above which are given by
\begin{equation}
V(\phi) = \frac{ \frac{C_o (1-A)}{2 a^m}    + \frac{1+\alpha -n}{1+\alpha} \;
\frac{ B_1}{ a^{3n}} }{ \left( \frac{ B_1}{a^{3n}} + \frac{C_o}{a^m} \right)^{\frac{\alpha}{1+\alpha}}},
\end{equation}
\begin{equation}
\dot{\phi}^2 = \frac{ \frac{n B_1}{(1+\alpha)a^{3n}} +
\frac{C_o(1+A)}{a^m} }{ \left(  \frac{B_1}{a^{3n}}+ \frac{C_o (1+A)}{a^m} \right)^{\frac{\alpha}{1+\alpha}}  } 
\end{equation}
The above equation reduces to that obtained in {\it Ref.} [26] for  $\alpha=1$, $A =0$ and $n=0$ and that obtained in {\it Ref.} [27] for  $n=0$.
Now we consider that the scalar field model of dark energy described by modified variable Chaplygin gas corresponds to holographic dark energy of the universe. In this section we reconstruct the correponding potential. 
Let us now consider a non-flat universe with $k \neq 0$ and use the holographic dark energy density given in (2) as
\begin{equation}
\rho_{\Lambda} = 3 c^2 M_P^2 L^{-2},
\end{equation}
where $L$ is the cosmological length scale for tracking the field
corresponding to holographic dark energy in the universe. The
parameter $L$ is defined as
\begin{equation}
L = a r (t).
\end{equation}
where $a(t)$ is the scale factor of the universe and $r(t)$ is relevant to the future event horizon of the universe. Using Robertson-Walker metric one gets [19]
\[
L = \frac{a (t)}{\sqrt{|k|}}  \;  sin \; \left[ \sqrt{|k|}
R_{h}(t)/a(t) \right] \;\; for \; \; \; k = +1 ,
\]
\[
\; \; \; =   R_h \;\; for \; \; k =0,
\]
\begin{equation}
 \; \; \; =  \frac{a (t)}{\sqrt{|k|}} \; \; sinh \; \left[
\sqrt{|k|} R_{h}(t)/a(t) \right]\; \; for \; \; k = - 1 .
\end{equation}
where $R_{h}$ represents the event horizon which is given by
\begin{equation}
R_h = a(t) \; \int_t^\infty \frac{dt'}{a(t')} = a(t) \; \int^{r_1}_o \frac{dr}{\sqrt{1 - k r^2}}.
\end{equation}
 Here $R_h$ is
 measured in $r$ direction and $L$ represents the radius of the event horizon measured on the sphere of the horizon. Using the definition of $\Omega_{\Lambda} = \frac{\rho_{\Lambda}}{\rho_{cr}} $ and $\rho_{cr} = 3 M_{P}^2 H^2$, one can derive  [20]
\begin{equation}
H L = \frac{c}{\sqrt{\Omega_{\Lambda}}}.
\end{equation}
Using eqs. (16)- (17), we determine the rate of change of $L$ with respect to
$ t$  which is
\[
\dot{ L} = \frac{c}{\sqrt{\Omega_{\Lambda}}} - \frac{1}{\sqrt{|k|}}
\; cos \; \left( \frac{\sqrt{|k|} \; R_h}{a(t)} \right) \; \; for \;
\; k = +1,
\]
\[
\; \; = \frac{c}{\sqrt{\Omega_{\Lambda}}} - 1 \; \; for \; \; k = 0,
\]
\begin{equation}
\; \; \; = \frac{c}{\sqrt{\Omega_{\Lambda}}} - \frac{1}{\sqrt{|k|}}
\; cosh \; \left( \frac{\sqrt{|k|} \; R_h}{a(t)} \right) \; \; for
\; \; k = - 1.
\end{equation}
Using eqs. (13) -(18) , we obtain the  holographic energy density $\rho_{\Lambda}$, which is given by
\begin{equation}
\frac{d\rho_{\Lambda}}{dt} =  - 2 H \left[ 1 -
\frac{\sqrt{\Omega_{\Lambda}}}{c} \; \frac{1}{\sqrt{|k|}} \; f(X)
\right]
 \; \rho_{\Lambda},
\end{equation}
here we use the notation, henceforth,
\begin{equation}
 f(X)  =  cos (X) \; \left[ 1, cosh (X) \right]  \; fo r\;   k =1 \; [0, -1],
\end{equation}
with $X = \frac{R_h}{a(t)}$.
 The energy conservation equation is
\begin{equation}
\frac{d\rho_{\Lambda}}{dt} + 3   H (1 + \omega_{\Lambda})  \rho_{\Lambda} = 0
\end{equation}
which is used to determine the equation of state parameter
\begin{equation}
\omega_{\Lambda}  =  - \left( \frac{1}{3} +  \frac{2
\sqrt{\Omega_{\Lambda}}}{3c} f(X) \right).
\end{equation}
Now  we assume  holographic dark energy density which is equivalent to  the modified   variable Chaplygin gas energy density.  The corresponding  energy density may be obtained  using eq.  (7).
The EOS parameter follows from eq. (1)
\begin{equation}
\omega  = \frac{p}{\rho} =  A  -   \frac{B(a) }{\rho^{\alpha +1}}.
\end{equation}
We now consider  correspondence between the holographic dark
energy and modified Chaplygin gas energy density. Using eqs. (7), (13) and (19), one obtains
\begin{equation}
B_o =  ( 3 c^2 M_P^2 L^{-2}) ^{\alpha +1} \; a^{3n} \; \left[ A + \frac{1}{3} +
\frac{2  \sqrt{\Omega_{\Lambda}}}{3  c}  f(X) \right],
\end{equation}
\begin{equation}
C_o =  ( 3 c^2 M_P^2 L^{-2})^{\alpha +1} \; a^m \; 
\left[ 1 - \frac{(1+\alpha) (A+ \frac{1}{3} + 2 \frac{
\sqrt{\Omega_{\Lambda}}}{3c}  \; f(X)}{(1+\alpha) (A + 1)  -n} \right].
\end{equation}
 Consequently one determines the scalar field potential which is given by
\begin{equation}
V( \phi)  = \frac{3 c^2 M_P^2 L^{-2}}{2} \left[
 1 - A + \frac{(1+\alpha)(1+A) -2n}{(1+\alpha)(1+A) -n} \; \left( A + \frac{1}{3} +\frac{2 \sqrt{\Omega_{\Lambda}  }   }{3} \; f(X) \right) \right],
\end{equation}
and the corresponding kinetic energy of the field is 
\begin{equation}
\dot{ \phi}^2  = 2  c^2 M_P^2 L^{-2} \left[ 1 -  \frac{
\sqrt{\Omega_{ \Lambda}}}{  c} f(X) \right].
\end{equation}
It is interesting to note that for $n = 0$  the potential reduces to the form that obtained by Paul {\it et al.} [27] and for $n=0$ and $A=0$, it  reduces to that form obtained by Setare [25] (where $B_o$ is to be replaced by $A$).
We now substitute $x$ $( = \ln a(t) )$, to transform  the time derivative to the derivative with logarithm of the  scale factor,  which is the most useful function in this case. Consequently from eq. (27) one obtains
\begin{equation}
\phi'= M_P \sqrt{ 2 \Omega_{\Lambda} \left( 1 - \frac{
\sqrt{\Omega_{\Lambda}}}{  c} f(X) \right)}
\end{equation}
where $()'$ prime represents derivative with respect to $x$. Thus, the evolution of the scalar field is given by
\begin{equation}
\phi (a) - \phi ( a_o) = \sqrt{2} M_P \int_{\ln a_o}^{\ln a} \sqrt{
\Omega_{\Lambda} \left( 1 -  \frac{ \sqrt{\Omega_{\Lambda}}}{  c}
f(X) \right)} \;  dx  .
\end{equation}

\section{Squared speed for  Holographic Dark Energy :}

We consider a closed  universe model ($k = 1$) in this case. The
dark energy equation of state parameter given by eq. (29) reduces to
\begin{equation}
\omega_{\Lambda}  = - \frac{1}{3} \left( 1 + \frac{2}{c}
\sqrt{\Omega_{\Lambda}} \; cos \; y \right)
\end{equation}
where  $y = \frac{R_H}{a(t)}$. The minimum value it can take is
 $\omega_{min} = - \frac{1}{3}  \left( 1 + 2 \sqrt{\Omega_{\Lambda}} \right)$ and one obtains
 a lower bound $\omega_{min} = - 0.9154$ for
 $\Omega_{\Lambda}= 0.76$ with $c = 1$.
Taking variation of the state parameter with respect to $x = \ln \; a$, we get [17]
\begin{equation}
\frac{\Omega_{\Lambda}'}{\Omega_{\Lambda}^2}= (1 - \Omega_{\Lambda})
\left( \frac{2}{c} \frac{1}{\Omega_{\Lambda}} cos \; y + \frac{1}{1
- a \gamma} \frac{1}{\Omega_{\Lambda}} \right)
\end{equation}
and the variation of equation of state parameter becomes
\begin{equation}
\omega_{\Lambda}' = - \frac{\sqrt{\Omega_{\Lambda}}}{3 c} \left[
\frac{1 - \Omega_{\Lambda}}{1 - \gamma a} + \frac{ 2
\sqrt{\Omega_{\Lambda}}}{c} \left(1 - \Omega_{\Lambda} cos^2
y\right) \right],
\end{equation}
where $\gamma = \frac{\Omega^{o}_k}{\Omega^{o}_m}$. We now introduce
the squared speed of holographic dark energy fluid  as
\begin{equation}
{\it v}_{\Lambda}^2 = \frac{dp_{\Lambda}}{d \rho_{\Lambda}} = \frac{\dot{p}_{\Lambda}}{\dot{\rho}_{\Lambda}} = \frac{p'_{\Lambda}}{\rho'_{\Lambda}},
\end{equation}
where  varaiation of eq. (30) w.r.t.  $x$ is  given by
\begin{equation}
p'_{\Lambda} = \omega'_{\Lambda} \rho_{\Lambda}+ \omega_{\Lambda} \rho'_{\Lambda}.
\end{equation}
Using the eqs. (41) and (42) we get
\[
{\it v}_{\Lambda}^2 = \omega'_{\Lambda} \frac{{\rho}_{\Lambda}}{{\rho'}_{\Lambda}} + \omega_{\Lambda}
\]
which now becomes
\begin{equation}
{\it v}_{\Lambda}^2  = - \frac{1}{3}  - \frac{2}{3 c} \sqrt{\Omega_{\Lambda} } \; cos y + \frac{1}{6 c} \; \sqrt{ \Omega_{\Lambda} } \left[ \frac{ \frac{1 - \Omega_{\Lambda} }{1 - \gamma a}+ \frac{2}{c} \sqrt{ \Omega_{\Lambda} } \left( 1 - \Omega_{\Lambda} \; cos^2 y \right) }{ 1 - \frac{\Omega_{\Lambda} }{c} \; cos y  } \right].
\end{equation}

\input{epsf}
\begin{figure}
\epsffile{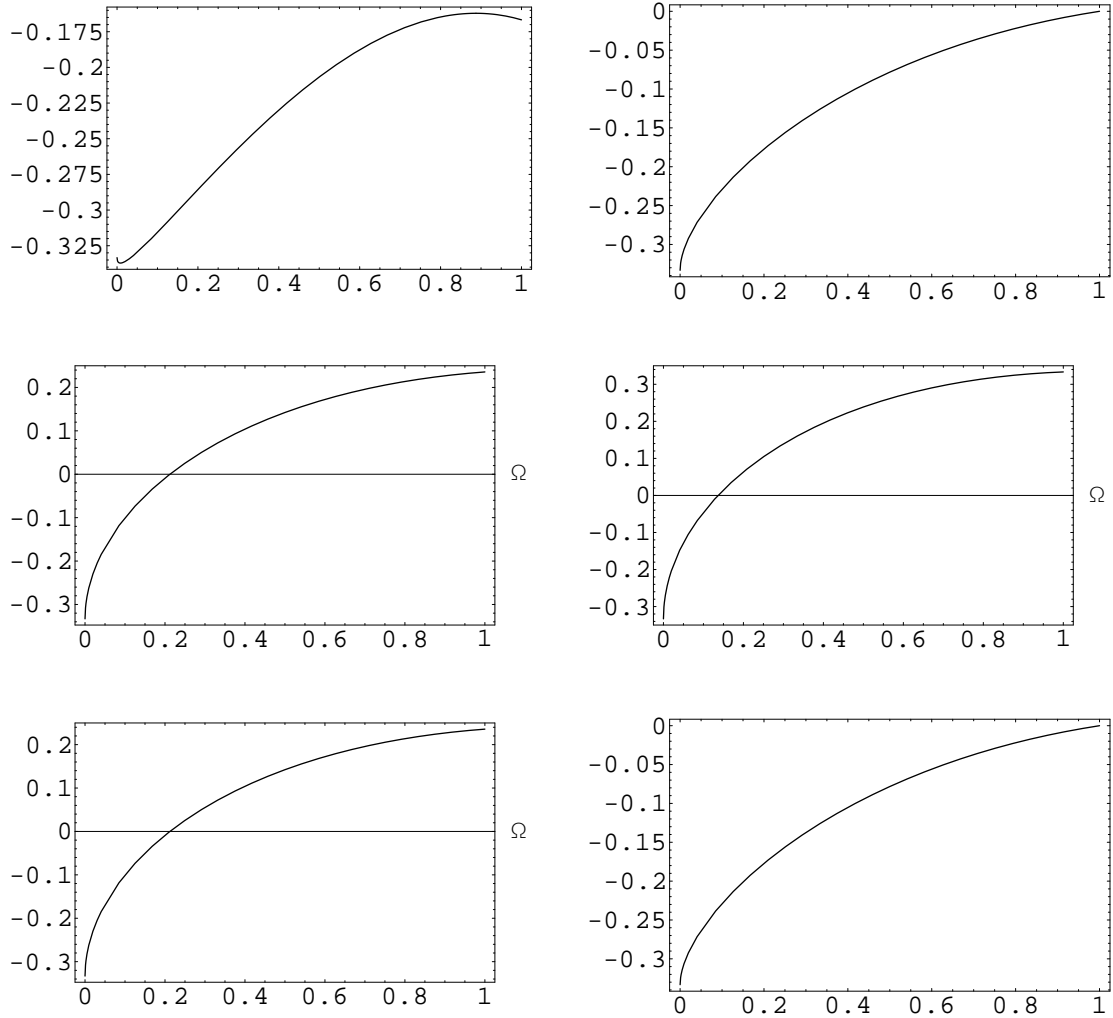}
\caption{shows the plot of ${\it v}_{\Lambda}^2$ versus $\Omega_{\Lambda} $ for different values of
$y$ with $c=1$, $\gamma = 1/3$ and $a=1$, in the first array the figures are  for $y =  \frac{\pi}{3} \; $ and $\; y =  \frac{\pi}{2} \; $, in the second array for  $ \; y =   \frac{1.5 \; \pi}{2} \; $,  $\; y =  \pi \; $ and in the third array for $y =   \frac{2.5 \; \pi}{2} \; $, $y =  \frac{3 \pi}{2}$.}
\end{figure}

The variation of ${\it v}_{\Lambda}^2$ with  $\Omega_{\Lambda}$ is shown in
 fig. 2 for different $y$ values. It is found that for a  given value of $c, \; a, \; \gamma$, the model admits a positive squared speed for $\Omega_{\Lambda} > 0$. However,
$\Omega_{\Lambda}$ is bounded below otherwise instability develops. We note also that for
 $  \frac{(2n+1) \pi}{2} < y < \frac{(2n+3) \pi}{2}$,
 (where $n$ is an integer)   no instability develops.
 We plot the case for $n = 0$ in  fig. 2, it is evident that for
 $y \leq \frac{\pi}{2}$ and $y \geq \frac{3 \pi}{2}$,  the squared speed
 for holographic dark energy  becomes negative which led to instability.
 But for the region $ \frac{\pi}{2} < y  <  \frac{3 \pi}{2}$ with $n=0$ no
 such instability develops. It is also found that for $y =0$ i.e., in flat case
 the holographic dark energy model is always unstable [28].

\pagebreak

\section{  Discussions : }

In this paper we explored  holographic dark energy model in a FRW
universe with a scalar field which describe the modified variable Chaplygin gas.  We consider correspondence of holographic dark energy and the  modified variable Chaplygin gas to reconstruct the potential. Since a complete theory of quantum gravity is yet to emerge, we adopt the above approach to understand the nature of dark energy. We determine the evolution of the field and reconstruct the potential of the Holographic dark energy in the case of flat, closed and open models of the universe. Although the cosmological observations predict a flat model of the universe, a closed universe with small positive curvature ($\Omega_k = 0.01$) is compatible with observations. So, in this paper we considered  non-flat case also. We give here a generalized expression for the potential and the kinetic energy term considering a modified variable Chaplygin gas [12,  13].  The holographic dark energy field and the corresponding potential  depend on  three parameters namely, $A$, $\alpha$ and $n$. The potential and the kinetic energy given by eqs. (11) and (12) reduce to that form obtained by Setare [26] for $A = 0$, $n=0$ and
$\alpha = 1$. However, the result obtained by Paul {\it et al.} [27] recovered for  $n=0$.
The stability of the holographic dark energy is studied in sec. 3 and found that the stability depends on the parameter  $\Omega_{\Lambda}$. The  evolution of the  holograpic dark energy field follows  samepattern in the modified Chaplygin gas, generalized Chaplygin gas and in the variable Chaplygin gas. However, the field potential is differs, which depends on the  EOS parameters $A = 0$, $n=0$ and
$\alpha = 1$. It is found that holographic field potential $V(\phi)$ become a constant for $n= \frac{(1+\alpha)(1+A)}{2}$. A free holographic field is permitted for the case $A=1$ and $n=1+\alpha$.

{\bf{ \it Acknowledgement :}}

Author would like to  thank IUCAA, Pune and IRC, Physics Department, North Bengal University (NBU) for providing facilities
to initiate the work. BCP is  thankful to the Third World Academy of
Sciences $\bf (TWAS)$, Italy for Associateship  to support a visit to the {\it
Institute of Theoretical Physics, Chinese Academy of Sciences,
Beijing } and University Grants Commission, New Delhi for financial support to carry out the work.

\pagebreak


\begin{thebibliography}{99}

\bibitem{kn:1}  A. G Riess {\it  et  al.},  {\it Astrophys. J.}  {\bf 607}, 665 (2004);
 S. Perlmutter { \it et al.,  Nature }  {\bf 51}, 391 (1998);
 S. Perlmutter { \it et  al., Astrophys.  J.}  {\bf 598}, 102 (2003);
 S. Perlmutter { \it et  al.,  Astrophys. J.}  {\bf 517}, 565 (1999); P. de Bernardis {\it et al.}  {\it Nature} {\bf 404} 955 (2000); M. Tegmark {\it et al.}, {\it Phys. Rev} {\bf D 69  }, 103501 (2004);  D. N. Spergel {\it et al.}, arXive ;  astro-ph/0603449. \\

\bibitem{kn:2} A. H. Guth,  {\it Phys. Rev.} {\bf D 23}, 347 (1981);  A. D. Linde, {\it Phys. Lett.} {\bf B 108},  389 (1982);  {\it Rep. Prog. Phys.} {\bf 47}, 925 (1984);
 R. H. Brandenberger, {\it Rev. Mod. Phys.}  {\bf 57}, 1 (1985);
 D. H. Lyth, D. Roberts and M Smith, {\it Phys. Rev.} {\bf D 57}, 7120 (1998);
  L. Mersini, {\it Mod. Phys. Lett.} {\bf A 16}, 1933 (2001);
  A. Linde,  { \it Int. J. Mod. Phys.} {\bf A17}, 89 (2002).

\bibitem{kn:3} A. A. Starobinsky, {\it Phys. Lett.} {\bf B 99}, 24 (1980).
\bibitem{kn:4} R. Fabri and M. D. Pollock, {\it Phys. Lett.} {\bf B 125}, 445 (1983); A. A. Starobinsky,  {\it Sov. Astron. Lett.} {\bf 9}, 302 (1983), {\it JETP Lett.} {\bf 42}, 152 (1986); S. W. Hawking and J. C. Luttrell, {\it Nucl. Phys.} {\bf B 247}, 250 (1984); L. A. Kofmann, A. D. Linde and A. A. Satrobinsky, {\it Phys. Lett.} {\bf B 157}, 361 (1985); A. Vilenkin, {\it Phys. Rev.} {\bf D 32}, 2511 (1985); M. B, Mijic, M. S. Morris and W. Suen, {\it Phys. Rev.} {\bf D 34}, 2934 (1986); S. Gottlober and V. Muller, {\it Class. Quantum Grav.}  {\bf 3}, 183 (1986); B. C. Paul, D. P. Datta and S. Mukherjee {\it Mod.  Phys. Lett.} {\bf A 3}, 843 (1988); G. Magnano and S. M. Sokolowski, {\it Phys. Rev.} {\bf D 50}, 5039 (1994); B. C. Paul and A. Saha, {\it Int. J. Mod. Phys.} {\bf D 11}, 493 (2002); S. Mukherjee, B. C. Paul, A. Beesham and S. D. Maharaj, arXive: gr-qc/0505103;  P.S. Debnath and B. C. Paul,  {\it Int. J. Mod. Phys.} {\bf D 15}, 189 (2006);
S. Mukherjee, B. C. Paul, N. K. Dadhich,  A. Beesham and S. D. Maharaj , {\it Class. Quantum Grav.}  {\bf 23}, 6927 (2006).
\bibitem{kn:5} A. Y. Kamenshchik, U. Moschella and V. Pasquier,
{\it  Phys. Lett.}  {\bf B 511}, 265 (2001); V. Gorini, A.
Kamenshchik, U. Moschella and V. Pasquier, arXive: gr-qc/0403062.

\bibitem{kn:6} V. Gorini,  A. Kamenshchik and U. Moschella, {\it Phys. Rev.} {\bf D 67} 063509 (2003); U. Alam, V. Sahni, T. D. Saini and A. A. Starobinsky, {\it Mon. Not. Roy. Astron. Soc.} {\bf 344}, 1057 (2003)

\bibitem{kn:7} M. C. Bento, O. Bertolami and A. A. Sen,
{\it Phys. Rev.} {\bf D 66}, 043507 (2002); V. Sahni, T. D. Saini,
A. A. Starobinsky and U. Alam, {\it JETP Lett.} {\bf 77}, 201
(2003).

\bibitem{kn:8} H. B. Benaoum, {\it Accelerated Universe from
Modified Chaplygin gas and tachyonic fluid}, arXive : hep-th/0205140.

\bibitem{kn:9} S. Nojiri and S. D. Odintsov,  {\it Phys. Rev.} {\bf D 70}, 103522 (2004), arXive: hep-th/0408170; {\it Phys. Rev.} {\bf D 72}, 023003 (2005), arXive: hep-th/0505215.

\bibitem{kn:10} J. D. Barrow, {\it Nucl. Phys.} {\bf B 310}, 743 (1988), {\it Phys. Lett.} {\bf B 235}, 40 (1990).

\bibitem{kn:11} V. Gorini,  A. Kamenshchik, U. Moschella and V. Pasquier, {\it Phys. Rev.} {\bf D
69 } 123512 (2004).

\bibitem{kn:12}  M. Jamil and M. A. Rashid, {\it Europhys. J } {\bf C 58}, 111 (2008).

\bibitem{kn:13}  Z-K. Guo and Y-Z. Zhang, {\it Phys. Lett. } {\bf B 645}, 326 (2007).

\bibitem{kn:14} G. Sethi, S. K. Singh, P. Kumar, D. Jain and A. Dev,{\it Int. J.  Mod.  Phys. } {\bf D 15},  1089 (2006); Z-K. Guo and Y-Z. Zhang, arXive: astro-ph/0509790.

\bibitem{kn:15} W. Fischler and L. Susskind, {\it Holography and Cosmology} arXive : hep-th/9806039; L. Susskind, {\it Holography in the flat space limit}, arXive : hep-th/9901079; D. Bigatti and L. Susskind, {\it TASI Lectures on the Holographic Principle},arXive : hep-th/0002044.

\bibitem{kn:16}  R. Bousso, {\it JHEP} {\bf 9907}, 004 (1999), {\it JHEP} {\bf 9906}, 028 ((1999); {\it Class. Quantum Grav},  {\bf 17}, 997 (2000).

\bibitem{kn:17}  S. D. H. Hsu, {\it Phys. Lett.} {\bf B 594}, 13 (2004).

\bibitem{kn:18}  M. Li, {\it Phys. Lett.} {\bf B 603}, 1 (2004).

\bibitem{kn:19} J. Zhang, X. Zhang and  H. Liu, {\it Phys. Lett.} {\bf B 651}, 84 (2007), arXive: 0706.1185; X. Zhang, {\it Phys. Lett.} {\bf B 648}, 1 (2007); M. R. Setare,{\it Phys. Lett.} {\bf B 653}, 116 (2007); N. Banerjee and D. Pavon, {\it Phys. Lett.} {\bf B 647}, 477 (2007); M. R. Setare, {\it JCAP}, {\bf 0701}, 023 (2007);  B. Chen, M. Li and Y. Wang, {\it Nucl. Phys.} {\bf B 774}, 256 (2007); M. R. Setare, {\it Phys. Lett.} {\bf B 644}, 99 (2007); M. R. Setare, {\it Eur. Phys. J.} {\bf C 50} 991 (2007);  M. R. Setare and S. Shafei, {\it JCAP} {\bf 0609}, 011 (2006);  M. R. Setare, {\it Phys. Lett.} {\bf B 642}, 1 (2006); J. P. Beltran Almeida and J. G. Pereirs, {\it Phys. Lett.} {\bf B 636}, 75 (2006), arXive : gr-qc/0602103;  X. Zhang, {\it Phys. Rev. } {\bf 74}, 103505 (2006), arXive : astro-ph/0609699;    S. Nojiri and S. D. Odintsov, {\it Gen. Rel. Grav.} {\bf 38}, 1285 (2006), hep-th/0506212; Y. Gong and Y.Z. Zhang, {\it Class. Quantum Grav.} {\bf 22}, 4895 (2005), hep-th/0505175; X. Zhang and F. Q. Wu, {\it Phys. Rev.} {\bf D 72}, 043524 (2005);  D. Pavon and W. Zimdahl, {\it AIP Conf. Proce.} {\bf 841}, 356 (2006), arXive: hep-th/0511053.

\bibitem{kn:20}  Q. G. Huang  and M. Li, {\it JCAP} {\bf 8}, 13 (2004), arXive:  astro-ph/0404229.
\bibitem{kn:21}  A. G. Cohen, D. B. Kaplan and A. E. Nelson, {\it Phys. Rev. Lett.} {\bf 82}, 4971 (1999).

\bibitem{kn:22}  P. Horava and  D. Minic, {\it Phys. Rev. Lett.} {\bf 85}, 1610 (2000); S. Thomas, {\it Phys. Rev. Lett.} {\bf 89}, 081301 (2002).

\bibitem{kn:23} F. Simpson, {\it JCAP} {\bf 0703}, 016 (2007).

\bibitem{kn:24} F. Leblond and A. W. Peet, {\it JHEP}, {\bf 0304},
048  (2003); N. Lambert, H. Liu and J. Maldacena, {\it JCAP} {\bf 0703}, 014 (2007) arXive :hep-th/0303139; C. J. Km. H. B. Kim. Y. B. Kim and O. K. Kwon, {\it
JHEP}, {\bf 0303}, 008 (2003); D. A. Steer and F. Vernizzi, {\it
Phys. Rev. }, {\bf D 70} 043527 (2004).

\bibitem{kn:25}  M. R.  Setare,{\it Phys. Lett. } {\bf B 654},  1 (2007).

\bibitem{kn:26} M. R. Setare, {\it Phys. Lett. } {\bf B 648}, 329 (2007).

\bibitem{kn:27} B. C. Paul, P. S. Thakur and S. Ghose, arXive:0809.3491.

\bibitem{kn:28} Y. S. Myung, {\it Phys. Lett. } {\bf B 652}, 223 (2007). 


\end{thebibliography}
\end{document}